\documentclass[floats,preprint,superscriptaddress]{revtex4}
\usepackage{bm}
\usepackage{epsfig}
\usepackage{rotating}

\newcommand{\nn}{\nonumber}
\newcommand{\beq}{\begin{equation}}
\newcommand{\eeq}{\end{equation}}
\newcommand{\bea}{\begin{eqnarray}}
\newcommand{\eea}{\end{eqnarray}}
\newcommand{\ben}{\begin{eqnarray*}}
\newcommand{\een}{\end{eqnarray*}}

\def\vec#1{{\bf #1}}
\def\D0{D\O}

\newcommand{\Asl}{\not\!\!A}

\newcommand{\vsl}{\not\!v}

\newcommand{\vk}{|\vec{k}_\gamma|}
\newcommand{\vpi}{|\vec{p}_{\pi^0}|}

\begin{document}

\title{Heavy-Quark Symmetry and the Electromagnetic Decays of Excited Charmed Strange Mesons}%

\author{Thomas Mehen}%
\email{mehen@phy.duke.edu}
\affiliation{Department of Physics, Duke University, Durham NC 27708, USA}
\affiliation{Jefferson Laboratory, 12000 Jefferson Ave., Newport News VA 23606}

\author{Roxanne P.~Springer}%
\email{rps@phy.duke.edu}
\affiliation{Department of Physics, Duke University, Durham NC 27708, USA}

\begin{abstract}
Heavy-hadron chiral perturbation theory (HH$\chi$PT) is applied to the decays of the  even-parity
charmed strange mesons, $D_{s0}(2317)$ and $D_{s1}(2460)$. Heavy-quark spin symmetry predicts the 
branching fractions for the three electromagnetic decays of these states to the ground states $D_s$ and $D_s^*$  in terms of a single
parameter.  The resulting predictions for two of the branching fractions are significantly higher than current upper limits from the
CLEO experiment. Leading  corrections to the branching ratios from chiral loop diagrams and spin-symmetry violating operators
in the HH$\chi$PT Lagrangian can  naturally account for this discrepancy. Finally the proposal that the  $D_{s0}(2317)$
$(D_{s1}(2460))$ is a hadronic bound state of a $D$ $(D^*)$ meson and a kaon is considered.  Leading order
predictions for electromagnetic branching ratios in this molecular scenario are in very poor agreement with existing data.
\end{abstract} 

\date{\today}
\maketitle

\vspace{0.33 in} 

\section{Introduction}

The discovery of the $D_{s0}(2317)$ \cite{Aubert:2003fg} and $D_{s1}(2460)$ \cite{Besson:2003cp} has revived interest in excited charmed
mesons. The dominant decay modes of these states are $D_{s0}(2317) \to D_s \pi^0$ and $D_{s1}(2460) \to D^*_s \pi^0$, with widths less than 7 
MeV~\cite{Besson:2003cp}. There is experimental evidence indicating that $D_{s0}(2317)$ and $D_{s1}(2460)$ are  $J^P=0^+$ and $1^+$
states, respectively~\cite{Krokovny:2003zq,Abe:2003jk}.  Had the masses of the $0^+$ and
$1^+$ states been above the threshold for the $S$-wave decay into $D$ mesons and  kaons, as anticipated in quark model \cite{Godfrey:xj,Godfrey:wj}
as well as lattice calculations \cite{Hein:2000qu,Boyle:1997rk,Lewis:2000sv}, they would have had widths of a few hundred MeV.  In
reality, the unexpectedly low masses make those decays kinematically impossible. The only available strong decay modes violate
isospin, accounting for the narrow widths.

The $D_{s0}(2317)$ and $D_{s1}(2460)$ can also decay electromagnetically. 
The possible decays are 
$D_{s1}(2460) \to D_s^* \gamma$, $D_{s1}(2460) \to D_s  \gamma$, and $D_{s0}(2317)\to D^*_s \gamma$.
The decay  $D_{s0}(2317)\to D_s \gamma$ is forbidden by parity conservation. In the heavy-quark limit, 
both the three electromagnetic decays and the  two strong decays
are related by heavy-quark spin symmetry~\cite{hqs}.  
Belle  has observed the decay $D_{s1}(2460) \to D_s\gamma$ from $D_{s1}(2460)$ produced in the decays
of $B$ mesons~\cite{Krokovny:2003zq} and from continuum $e^+e^-$ production~\cite{Abe:2003jk}. 
The ratio of the electromagnetic branching fraction  to the isospin violating one pion decay 
reported by the experiment is
\bea\label{belle_one}
\frac{ {\rm Br}(D_{s1}(2460)\to D_s \gamma)}{{\rm Br}(D_{s1}(2460)\to D^*_s \pi^0)} = \left\{
\begin{array}{lc} 
0.38 \pm 0.11 \pm 0.04 & \qquad  [3]  \\
0.55 \pm 0.13 \pm 0.08 & \qquad  [4] 
\end{array} \right. \,.
\eea 
In each case the first error is statistical and the second systematic.
 The other electromagnetic decays have not been observed. CLEO quotes
the following bounds on the branching fraction ratios~\cite{Besson:2003cp}:
\bea \label{cleo}
\frac{ {\rm Br}(D_{s1}(2460)\to D^*_s \gamma)}{{\rm Br}(D_{s1}(2460)\to D^*_s \pi^0)} < 0.16 \quad
\frac{ {\rm Br}(D_{s0}(2317)\to D^*_s \gamma)}{{\rm Br}(D_{s0}(2317)\to D_s \pi^0)} < 0.059 \,.
\eea 
(The BELLE collaboration  quotes weaker lower bounds of 0.31 and 0.18, respectively, for these ratios~\cite{Abe:2003jk}.)

In this paper the decays of the $D_{s0}(2317)$ and $D_{s1}(2460)$ are analyzed using heavy-hadron chiral
perturbation theory (HH$\chi$PT)~\cite{hhchpt}. HH$\chi$PT is an effective theory applicable to the low energy  strong and
electromagnetic interactions of particles containing a heavy quark. It incorporates the approximate  heavy-quark and chiral symmetries
of QCD. Corrections to leading order  predictions can be computed in an expansion in $\Lambda_{\rm QCD}/m_Q$, $M/\Lambda_\chi$, and 
$p/\Lambda_\chi$, where $m_Q$ is the heavy quark mass, $M$ is a Goldstone boson mass, $p$ is the typical
momentum in the decay, and $\Lambda_\chi$ is the chiral symmetry breaking scale.

In section II,   the leading order HH$\chi$PT predictions for the branching ratios are derived:
\bea\label{predict}
\frac{ {\rm Br}(D_{s1}(2460)\to D^*_s \gamma)}{{\rm Br}(D_{s1}(2460)\to D^*_s \pi^0)} = 0.37 \pm 0.07 \quad
\frac{ {\rm Br}(D_{s0}(2317)\to D^*_s \gamma)}{{\rm Br}(D_{s0}(2317)\to D_s \pi^0)} = 0.13 \pm 0.03 \, .
\eea
(Leading order calculations of strong and electromagnetic decays were first done in
Refs.~\cite{Colangelo:1993zq,Bardeen:2003kt,Colangelo:2003vg}.)  These predictions deviate significantly from the CLEO limits. 
At next-to-leading order (NLO) there are
$O(1/m_Q)$ suppressed heavy-quark spin-symmetry violating operators as well as one-loop chiral corrections to the
electromagnetic decays.  Once these effects are included, predictions for the ratios
in Eq.~(\ref{cleo}) can be made consistent with the present  experimental bounds with coupling constants in the
Lagrangian  of natural size. 

The splitting between the even- and odd-parity doublets should be approximately  the same for both bottom strange and charmed strange
mesons.  Therefore it is likely that the $B_s$ even-parity states will be below threshold for decay into kaons and  narrow
like their charm counterparts. The calculations of this paper can also be applied to the electromagnetic and strong decays of
even-parity $B_s$ mesons when these states are discovered.

The leading order HH$\chi$PT Lagrangian used in this paper is invariant under nonlinearly realized chiral 
$SU(3)_L \times SU(3)_R$ and no further assumptions are made about the mechanism of chiral symmetry breaking. Models
that treat the  $D_{s0}(2317)$  and $D_{s1}(2460)$ as the $0^+$ and $1^+$ chiral partners of the ground state charm
strange mesons are proposed in  Refs.~\cite{Bardeen:2003kt,Nowak:2003ra,Bardeen:1993ae,Nowak:1992um}.  In these
models, referred to as parity-doubling models,  the even-parity and odd-parity mesons are placed in a
linear representation of chiral $SU(3)_L \times SU(3)_R$. These fields couple in a chirally  invariant manner to a
field $\Sigma$ that transforms in the $(3,\bar 3)$ of $SU(3)_L \times SU(3)_R$. The $\Sigma$ field  develops a
vacuum expectation value that breaks the chiral symmetry. The resulting nonlinear sigma model of Goldstone bosons
coupled  to heavy mesons has the same operators as the HH$\chi$PT Lagrangian used in this paper.  The assumed
mechanism of chiral symmetry breaking in parity doubling models predicts relationships between coupling constants in
the HH$\chi$PT Lagrangian. For example, the parity doubling models predict that the  hyperfine splittings of the
even- and odd-parity doublets are equal. This is in agreement with experimental observations. Other relationships
between coupling constants in HH$\chi$PT are predicted~\cite{Beane:2002td,Beane:1995yw} by the theory  of algebraic
realizations of chiral symmetry~\cite{Weinberg:xn}, in which hadrons are placed in reducible representations of
$SU(3)_L \times SU(3)_R$. QCD sum rules have also been used to calculate some of the HH$\chi$PT
couplings~\cite{Colangelo:1995ph}. When more data on the electromagnetic  decays of even-parity $D_s$ and $B_s$
mesons becomes available, the formulae derived in this paper can be used to extract the relevant couplings and test
these theories. 

The low mass of the $D_{s0}(2317)$ and $D_{s1}(2460)$ has prompted reexamination of quark models~\cite{Cahn:2003cw,Godfrey:2003kg,
Lucha:2003gs,:2003dp,Azimov:2004xk} as well as speculation that these states are exotic. Possibilities include $D K$ molecules 
\cite{Barnes:2003dj,Nussinov:2003uj,Chen:2004dy}, $D_s \pi$ molecules \cite{Szczepaniak:2003vy}, and tetraquarks
\cite{Nussinov:2003uj,Terasaki:2003qa,Terasaki:2003dj,Terasaki:2004yx, Vijande:2003wk,Cheng:2003kg}. Masses have been calculated in
lattice QCD~\cite{Bali:2003jv,Dougall:2003hv,diPierro:2003iw}, heavy-quark effective theory (HQET) sum rules~\cite{Dai:2004yk,Dai:2003yg},
and potential as well as other  models~\cite{Cahn:2003cw,Lucha:2003gs,:2003dp,Azimov:2004xk,Deandrea:2003gb,Hsieh:2003xj}.  The results of
some of these papers are contradictory. For example, the lattice  calculation of Ref.~\cite{Bali:2003jv} yields a $0^+$-$0^-$ mass
splitting about 120 MeV greater than experimentally observed, quotes errors of about 50 MeV, and argues this is evidence for an exotic
interpretation of the state. On the other hand,  the lattice calculation of Ref.~\cite{Dougall:2003hv}  obtains similar numerical results
but concludes that uncertainties in the  calculation are large enough to be consistent with a conventional $c\bar{s}$ $P$-wave state. Some
quark model analyses ~\cite{Deandrea:2003gb,Azimov:2004xk,:2003dp} conclude that  interpreting the states as conventional $c\bar s$ $P$-wave
mesons  naturally fits the observed data, others reach the opposite conclusion~\cite{Cahn:2003cw,Hsieh:2003xj}. 

There have been some attempts to  determine the nature of the $D_{s0}(2317)$ and $D_{s1}(2460)$ from the observed pattern of
decays~\cite{Godfrey:2003kg} as well as their production in $b$-hadron decays~\cite{Chen:2003jp,Datta:2003re,Datta:2003yk,Huang:2004et}.
Ref.~\cite{Godfrey:2003kg} argues that the total width and electromagnetic branching ratios can distinguish between $c\bar{s}$ $P$-wave
states and $D K$ molecules, and gives predictions for these branching ratios calculated in the quark model.  
Refs.~\cite{Chen:2003jp,Datta:2003re} argue that the observed branching fractions for $B \to D_{s0}(2317) D^{(*)}$ and $B \to D_{s1}(2460)
D^{(*)}$ are smaller then expected for $c\bar{s}$ $P$-wave states, suggesting that these states are exotic. These analyses assume an
unproven (but plausible) factorization conjecture for the decays as well as quark model  estimates for the $D_{s0}(2317)$ and $D_{s1}(2460)$
decay constants, and have recently been extended to $\Lambda_b$~\cite{Datta:2003yk} and semileptonic $B_s$ decays~\cite{Huang:2004et}.

Section III addresses the question of whether a model independent analysis of the decays can provide insight into the nature of the
$D_{s0}(2317)$ and $D_{s1}(2460)$. In HH$\chi$PT the fields describing the $0^+$ and $1^+$ mesons are added to the Lagrangian by hand. The
only assumption made about these states is that the light degrees of freedom in the hadron are in the $\bar{3}$ of $SU(3)$ and have
$j^p=\frac{1}{2}^+$. (In this paper, $J^P$ refers to the angular momentum and parity of a heavy meson, and  $j^p$ to the angular momentum
and parity of the light degrees of freedom.) Light degrees of freedom with these quantum numbers could be an $\bar s$ quark in an orbital
$P$-wave or $\bar s q \bar q$ quarks  all in an $S$-wave. Therefore, a conventional quark model $c\bar{s}$ $P$-wave state and an
unconventional $c \bar s \bar q q$ tetraquark state will be represented by fields having the same transformation properties in the
HH$\chi$PT  Lagrangian. HH$\chi$PT predictions for the ratios in Eqs.~(\ref{belle_one}-\ref{cleo}) are valid for either interpretation, and
so cannot distinguish between these two scenarios. 

However, if $D_{s0}(2317)$ ($D_{s1}(2460)$) is modeled as a bound state of a $D$ ($D^*$) meson and a kaon the
predictions for the electromagnetic branching ratios will be different. In this scenario, instead of adding the
even-parity heavy-quark doublet to the Lagrangian by hand, the dynamics of the theory  containing only the ground
state heavy-quark doublet and Goldstone bosons generate the observed $D_{s0}(2317)$ and  $D_{s1}(2460)$. This
interpretation has been pursued in
Refs.~\cite{vanBeveren:2003kd,vanBeveren:2003hj,Kolomeitsev:2003ac,Hofmann:2003je,Lutz:2004tv}. In this scenario
the binding energy is only about 40 MeV, so the mesons in the hadronic bound state are nonrelativistic. The decay
rates can be calculated by convolving the  unknown nonrelativistic wavefunction with leading order HH$\chi$PT
amplitudes for $D^{(*)} K \to D_s^{(*)} \gamma$ and  $D^{(*)}K \to D_s^{(*)} \pi^0$. Dependence on the bound state
wavefunction drops out of the ratios in Eqs.~(\ref{belle_one}-\ref{cleo}).  The resulting predictions for these
branching ratios are much larger than experiment. Furthermore, the branching ratio for $D_{s1}(2460) \to D_s \gamma$
is predicted to be the smallest of the three, in direct conflict with experimental observations. A
$D K$ molecular interpretation of the  $D_{s0}(2317)$ and $D_{s1}(2460)$ is disfavored by the existing data on
electromagnetic branching fractions.

\section{Electromagnetic and Strong Decays in HH$\chi$PT}

In the heavy-quark limit, hadrons containing a single heavy quark fall into doublets of the $SU(2)$ heavy-quark spin symmetry group. 
Heavy hadrons can be classified by the total angular momentum and parity quantum numbers of their light degrees of freedom, $j^p$. The ground 
state doublet has $j^p =\frac{1}{2}^-$ and therefore the mesons in the doublet are $0^-$ and $1^-$ states. In HH$\chi$PT, these states 
are combined into a single field~\cite{hhchpt}
\begin{eqnarray}\label{H}
H_a = \frac{1+ \vsl}{2}\left(H^\mu_a \gamma_\mu - H_a \gamma_5 \right) \, ,
\end{eqnarray}
where $a$ is an $SU(3)$ index. In the charm sector, $H_a$ consists of the $(D^0,D^+,D_s^+) \sim (c\bar u, c \bar d, c \bar s)$ 
pseudoscalar mesons and 
$H^\mu_a$ are the $(D^{0*},D^{+*},D_s^{+*})$ vector mesons. The doublet with light degrees of freedom
$j^p=\frac{1}{2}^+$ consists of mesons whose quantum numbers are $0^+$ and $1^+$. These are combined into 
the field~\cite{Falk:1991nq}
\begin{eqnarray}\label{S}
S_a = \frac{1+ \vsl}{2}\left(S^\mu_a \gamma_\mu \gamma_5 - S_a \right) \, ,
\end{eqnarray}
where the scalar states in the charm sector are $S_a = D_{0a}$ and the axial vectors
are $S^\mu_a =D_{1 a}$. 

The relevant strong interaction terms in the HH$\chi$PT chiral Lagrangian are~\cite{hhchpt,Falk:1992cx}
\begin{eqnarray}\label{Lag}
{\cal L} &=& \frac{f^2}{8}{\rm Tr}[\partial_\mu  \Sigma \,   \partial^\mu \Sigma^\dagger] 
+ \frac{ f^2 B_0}{4}{\rm Tr}[m_q \Sigma + \Sigma^\dagger m_q] \nn\\
&& -{\rm Tr}[\overline{H}_a  i v\cdot D_{b a}  H_b]
+  {\rm Tr}[\overline S_a (i v\cdot D_{b a}-\delta_{SH} \,\delta_{ab}) S_b] \nn  \\
&& + \, g \, {\rm Tr}[ \overline H_a H_b \Asl_{ba} \gamma_5] + g^\prime \, {\rm Tr}[ \overline S_a S_b \Asl_{b a} \gamma_5]
+ h ( {\rm Tr}[ \overline H_a S_b \Asl_{b a} \gamma_5] + {\rm h.c.} ) \nn \\
&&- \frac{\Delta_H}{8} {\rm Tr}[\overline{H}_a \sigma^{\mu \nu} H_a \sigma_{\mu \nu}]
+ \frac{\Delta_S}{8}  {\rm Tr}[\overline{S}_a \sigma^{\mu \nu} S_a \sigma_{\mu \nu}] \, .
\end{eqnarray} 
The first line in Eq.~(\ref{Lag}) is the leading order chiral Lagrangian for the octet of Goldstone bosons. $f$
is the octet meson decay constant. The conventions for
defining $\Sigma$ in terms of meson fields,  the chiral covariant derivative, $D_{a b}$, and the axial vector vector 
field, $A^\mu_{a b}$, are identical to those of Ref.~\cite{Stewart:1998ke}.  Here $m_q$ is the light quark mass matrix. 
The second line of Eq.~(\ref{Lag}) contains the
kinetic terms for the fields $H_a$ and $S_a$ and the couplings to two and more pions determined by chiral symmetry.
The  parameter $\delta_{SH}$ is the residual mass of the $S_a$ field. The $H_a$ residual mass can be set to zero by
an appropriate definition of the $H_a$ field, and this convention is adopted here. Then
$\delta_{SH}$ is the difference between the spin-averaged masses of the even- and odd-parity doublets in the heavy
quark limit.  The third line contains the couplings of  $H_a$ and $S_a$ to the axial vector field $A^\mu_{a b} = -
\partial^\mu \pi_{a b} /f + ...$. These terms are responsible for transitions involving a single pion.
The couplings $g$, $g^\prime$, and $h$ are parameters that are not determined by the HH$\chi$PT symmetries. The last line
in Eq.~(\ref{Lag}) contains operators that give rise to $1^-$-$0^-$ and $1^+$-$0^+$ hyperfine splittings, which are $\Delta_H$ and
$\Delta_S$, respectively. Since the splittings should vanish in the heavy quark limit, $\Delta_S \sim \Delta_H \sim
\Lambda_{\rm QCD}^2/m_Q$. The parameters $\Delta_S$ and $\Delta_H$ are independent in HH$\chi$PT so there is no relation between the
hyperfine splitting in the even- and odd-parity doublets. In parity doubling models, $\Delta_H = \Delta_S$ at tree
level, in agreement with the observation that  hyperfine splittings are equal to within 2 MeV.

Electromagnetic effects are incorporated by gauging the $U(1)_{\rm em}$ subgroup of 
$SU(3)_L \times SU(3)_R$ and adding terms to the Lagrangian involving the gauge invariant field
strength, $F_{\mu \nu}$. Gauging derivatives in Eq.~(\ref{Lag}) does not yield terms which
can mediate the $(0^+,1^+) \to(0^-,1^-)$ electromagnetic decays at tree level. The leading contribution to these decays comes from the 
operator
\begin{eqnarray}\label{emlo}
{\cal L} = \frac{e \tilde \beta}{4} {\rm Tr} [ \overline{H}_a S_b \sigma^{\mu \nu}] F_{\mu \nu} Q^\xi_{b a} \, ,  
\end{eqnarray}
where $Q^\xi_{b a} = \frac{1}{2}(\xi Q \xi^\dagger + \xi^\dagger Q \xi)_{b a}$, $\xi^2 = \Sigma$, and $Q  = {\rm diag}(2/3,-1/3,-1/3)$ 
is the light quark electric charge matrix.  
A tree level calculation of the decay rates using Eq.~(\ref{emlo})  shows that 
\begin{eqnarray}\label{lo}
\Gamma[1^+_a \to 1^-_a \gamma] &=& \frac{2}{3}\alpha e_q^2 \tilde \beta^2 \frac{m_{1^-_a}}{m_{1^+_a}}\vk^3 \nn \\
\Gamma[1^+_a \to 0^-_a \gamma] &=& \frac{1}{3}\alpha e_q^2 \tilde \beta^2  \frac{m_{0^-_a}}{m_{1^+_a}}\vk^3 \nn \\
\Gamma[0^+_a \to 1^-_a  \gamma] &=& \alpha e_q^2 \tilde \beta^2 \frac{m_{1^-_a}}{m_{0^+_a}}\vk^3 \, ,
\end{eqnarray}
where $e_q$ is the electric charge of the light valence quark, $\alpha$ is the fine-structure constant, and
$\tilde \beta$ is the unknown parameter in Eq.~(\ref{emlo}). The three-momentum of the photon in the decay is $\bf k_\gamma$  and $m_{J^P_a}$
is the mass of the heavy meson with quantum numbers $J^P_a$. In the heavy-quark limit, the members of each doublet are degenerate and the
phase space is  the same  for all three decays. If differences in phase space are neglected the decay rate ratios are   $\Gamma[1^+_a \to
1^-_a \gamma]:\Gamma[1^+_a \to 0^-_a \gamma]:\Gamma[0^+_a \to 1^-_a  \gamma]=2:1:3$. Differences in the phase space factors are formally 
$O(1/m_Q)$  but in practice it is critical to include these effects to make sensible predictions. For the charmed strange mesons using
the physical masses gives
\begin{eqnarray}
\Gamma[D_{s1}(2460) \to D_s^* \gamma] &=& (\tilde \beta \,  {\rm GeV})^2 \, 15.6 \,{\rm keV} \nn \\
\Gamma[D_{s1}(2460) \to D_s \gamma] &=& (\tilde \beta \, {\rm GeV})^2 \, 18.7 \, {\rm keV} \nn \\
\Gamma[D_{s0}(2317) \to D^*_s \gamma] &=&(\tilde \beta \,  {\rm GeV})^2 \, 5.6 \,{\rm keV} \, . 
\end{eqnarray}
The rates are then in the following ratios:
\begin{eqnarray}\label{ratio}
\Gamma[D_{s1}(2460) \to D_s^* \gamma]:\Gamma[D_{s1}(2460) \to D_s \gamma]: 
\Gamma[D_{s0}(2317) \to D^*_s \gamma] =0.83:1.0:0.30\, .
\end{eqnarray}
Note that the rate for $D_{s1}(2460) \rightarrow D_s \gamma$,  smallest in the exact heavy-quark limit,
is actually the largest when phase space effects are included since $\vk$ is largest for this decay. 

To compare with the ratio measured by Belle, the isospin violating strong decays must be calculated.
These decays proceed through $\eta -\pi^0$ mixing. The result is~\cite{Colangelo:2003vg}  
\bea\label{singpi}
\Gamma[D_{s1}(2460)\to D^*_s \pi^0] 
&=& \frac{h^2 \theta^2}{3 \pi f^2} \frac{m_{D^*_s}}{m_{D_{s1}(2460)}} E^2_{\pi^0} |{\bf p}_{\pi^0}| \nn  \\
&=&  h^2 \, \left\{\begin{array}{c}  
17.0 \, {\rm keV} \qquad \,\,\, {\rm if} \quad f=f_\pi = 130 \, {\rm MeV} \\
9.8 \, {\rm keV} \qquad {\rm if} \quad f=f_\eta = 171 \, {\rm MeV} 
\end{array} \right. \nn \\ 
\Gamma[D_{s0}(2317)\to D_s \pi^0] 
&=&\frac{h^2 \theta^2}{3 \pi f^2} \frac{m_{D_s}}{m_{D_{s0}(2317)}} E^2_{\pi^0}  |{\bf p}_{\pi^0}| \nn  \\
&=&  h^2 \, \left\{\begin{array}{c}  
16.9 \, {\rm keV} \qquad \,\,\,{\rm if} \quad f=f_\pi = 130 \, {\rm MeV} \\
9.8 \, {\rm keV} \qquad {\rm if} \quad f=f_\eta = 171 \, {\rm MeV} 
\end{array} \right.   \,,
\eea
where $\theta =\sqrt{3}/2 (m_d-m_u)/(2 m_s -m_d -m_u)=0.01$ is the $\eta-\pi^0$ mixing angle.  
$E_{\pi^0}$ and ${\bf p}_{\pi^0}$ are the energy and three-momentum of the $\pi^0$, respectively.
At tree level $f= f_\pi = f_\eta$. The difference between the two predictions provides 
an estimate of the uncertainty due to higher order $SU(3)$ violating effects.

The branching fraction ratios in Eqs.~(\ref{belle_one}-\ref{cleo}) depend only on the  ratio $\tilde \beta^2/h^2$ at leading order in
HH$\chi$PT. To obtain $h^2$ separately, a measurement  of an excited strong decay width is needed. Currently $h^2$ cannot  be extracted
from the strange sector because only loose experimental bounds on $\Gamma[D_{s0}(2317)]$ and  $\Gamma[D_{s1}(2460)]$ exist.  Until
measurements of these widths are dramatically improved, $h^2$ can be estimated using data on nonstrange even-parity $D$ meson widths.
CLEO~\cite{Anderson:1999wn} has observed preliminary evidence for  the  $D^0_1$ $(J^P =1^+)$ meson. (Here the superscript refers to the
particle charge.) More recently, Belle~\cite{Abe:2003zm} has reported observing  even-parity  $D_0^0$ $(J^P =0^+)$ and $D_1^0$ states.
Finally, FOCUS~\cite{Link:2003bd} has observed broad structures  in excess  of background in the $D^+ \pi^-$ and $D^0 \pi^+$ invariant
mass spectra. FOCUS does not claim to  observe an excited charm resonance but does fit the excess with a Breit-Wigner to determine the
resonance  properties required to explain their data. The masses and widths reported by all three experiments are collected  in
Table~\ref{tab0}. The experiments all quote several errors which have been combined in  quadrature for simplicity.
\begin{table}[!h]
\begin{center} \begin{tabular}{ccc|cccccc} 
&  Experiment && Particle($J^P$) && Mass (MeV) && Width (MeV)&\\
\hline
& CLEO \cite{Anderson:1999wn}  && $D^0_1(1^+)$   && $2461^{+53}_{-48}$ && $290^{+110}_{-91}$&\\
\hline
& Belle \cite{Abe:2003zm}  && $D^0_0(0^+)$   && $2308\pm 36$ && $276\pm 66$&\\
&                               && $D^0_1(1^+)$   && $2427\pm 36$ && $384^{+130}_{-105}$&\\
\hline
& FOCUS \cite{Link:2003bd}  && $D^0_0(0^+)$   && $2407\pm 41$ && $240\pm 81$&\\
&                               && $D^+_0(0^+)$   && $2403\pm 38$ && $283\pm 42$&\\
\end{tabular} \end{center}
{\tighten  \caption{Masses and widths of even-parity non-strange charmed mesons, $D_J^Q$,
where $Q$ is the electric charge.}
\label{tab0} }
\end{table}
Note that the CLEO and Belle measurements of the $D^0_1$ are consistent with each other 
while the central value of the $D^0_0$ mass obtained by FOCUS is 99 MeV higher than the 
central value of the Belle measurement.  
Furthermore, the FOCUS $D^0_0$ mass is actually greater than the mass of the $D_{s0}(2317)$. If the effect
observed by FOCUS is a scalar $D$ resonance, it seems implausible that this resonance 
is related to the $D_{s0}(2317)$ by $SU(3)$ symmetry. Therefore, the FOCUS data will not be used to estimate $h^2$.
Even the masses obtained by CLEO and Belle are large compared to expectations based on $SU(3)$ 
symmetry. Combining the strange sector $0^+$-$0^-$ and $1^+$-$1^-$ mass splittings with  $SU(3)$ symmetry 
leads to the prediction that  the $D^0_0$ mass is 2212  MeV and the $D^0_1$ mass is 2355 MeV~\cite{Bardeen:2003kt}.

Applying the leading order expression for the decay widths of the nonstrange $0^+$ and $1^+$ mesons
\bea\label{decay}
\Gamma[D_1^{0}] &=& \frac{h^2}{2 \pi f^2}\left( \frac{m_{D^{+*}}}{m_{D_1^{0}}} E^2_{\pi^-} |{\bf p}_{\pi^-}|
+ \frac{1}{2}\frac{m_{D^{0*}}}{m_{D_1^{0}}} E^2_{\pi^0}  |{\bf p}_{\pi^0}| \right) \nn \\
\Gamma[D_0^{0}] &=& \frac{h^2}{2 \pi f^2}\left( \frac{m_{D^+}}{m_{D_0^{0}}} E^2_{\pi^-} |{\bf p}_{\pi^-}| 
+ \frac{1}{2}\frac{m_{D^0}}{m_{D_0^{0}}} E^2_{\pi^0}  |{\bf p}_{\pi^0}| \right) \, ,
\eea
to the CLEO and Belle data yields $h^2 = 0.39 \pm 0.13$ from the $0^+$ decays and  $h^2 = 0.49 \pm 0.14$
from the $1^+$ decay. The error in each case is obtained by adding  in quadrature the uncertainty in the decay rate 
from varying the mass within the allowed range and the experimental error in the decay rate. 
If the two results are averaged  $h^2 = 0.44 \pm 0.11$.  
This estimate of $h^2$ is consistent with the bound $h^2 \leq  0.86$ extracted from an Adler-Weisberger type sum
rule for $\pi B$  scattering~\cite{Chow:1995ca}.  (To obtain this bound $g=0.27$~\cite{Stewart:1998ke} is used in the result of
Ref.~\cite{Chow:1995ca}.) It is also consistent with a calculation of  $h=-0.52 \pm 0.17 $ obtained
using QCD sum rules in Ref.~\cite{Colangelo:1995ph}.

The lowest order HH$\chi$PT prediction for the ratio measured by the Belle collaboration is 
\bea 
\frac{ {\rm Br}(D_{s1}(2460)\to D_s \gamma)}{{\rm Br}(D_{s1}(2460)\to D^*_s \pi^0)} 
=  \left( \frac{\tilde \beta \, {\rm GeV}}{h}\right)^2  \times 
\left\{\begin{array}{c}  
1.1 \qquad {\rm if} \quad f=f_\pi = 130 \, {\rm MeV} \\
1.9 \qquad {\rm if} \quad f=f_\eta = 171 \, {\rm MeV} 
\end{array} \right.
\, .
\eea
Averaging the results of the two Belle measurements, $(\tilde \beta \, {\rm GeV}/h)^2 = 0.40 \pm 0.08 \,(0.23 \pm 0.05)$ or 
$ |\tilde \beta| = 0.42 \pm 0.07 \,(0.32 \pm 0.05) \, {\rm GeV}^{-1}$, where 
$f=f_\pi \, (f =f_\eta)$. The error is estimated by first combining the statistical and systematic errors in quadrature
for each measurement, then combining the two measurements assuming they are independent.
 The extracted values for $\tilde \beta$ and $h$
are consistent with expectations based on naturalness. Plugging the value of $(\tilde \beta \, {\rm GeV}/h)^2$
into expressions for the unobserved electromagnetic decays yields the following 
predictions for the branching fraction ratios:
\bea
\frac{ {\rm Br}(D_{s1}(2460)\to D^*_s \gamma)}{{\rm Br}(D_{s1}(2460)\to D^*_s \pi^0)} = 0.37 \pm 0.07 \quad
\frac{ {\rm Br}(D_{s0}(2317)\to D^*_s \gamma)}{{\rm Br}(D_{s0}(2317)\to D_s \pi^0)} = 0.13 \pm 0.03 \, .
\eea
Both predictions are in excess of bounds quoted by the CLEO experiment.
Heavy-quark spin symmetry predicts branching ratios for the electromagnetic
decays $D_{s1}(2460)\to D^*_s \gamma$  and $D_{s0}(2317)\to D^*_s \gamma$ that are more than a factor of two in excess of the experimental
upper limits.

In the rest of this section the leading corrections 
to both electromagnetic and strong decays are analyzed. Because $\Lambda_{\rm QCD}/m_c \sim 1/3$
corrections to heavy-quark spin symmetry predictions can be rather large for charm hadrons. These corrections can be 
systematically analyzed using HH$\chi$PT. For example, the pattern of deviations from heavy-quark 
spin symmetry predictions for the one-pion decays of excited D-wave charm mesons \cite{Lu:px} can be 
understood by analyzing the structure of spin-symmetry violating operators
appearing at $O(1/m_c)$ in the HH$\chi$PT Lagrangian \cite{Falk:1995th}. A ratio of decay widths for which the $O(1/m_c)$ correction
vanishes agrees well with data. There is another ratio for which the leading order heavy-quark spin symmetry prediction fails
rather badly. In this case the leading $O(1/m_c)$ correction is multiplied by a large
numerical coefficient. Thus  HH$\chi$PT is a useful tool for determining the robustness of 
predictions based on heavy-quark spin symmetry.

Spin-symmetry violating operators that contribute  to $S \to H$ transitions must have the Dirac structure  ${\rm Tr} [
\overline{H}\sigma^{\mu \nu} S \gamma_5]$ or ${\rm Tr} [ \overline{H}\sigma^{\mu \nu} S \gamma^\alpha]$. Operators with  $\overline{H} S$
conserve spin symmetry, while operators with $\overline{H}\sigma^{\mu \nu} S$ violate spin symmetry. Operators with $\overline{H}
\gamma^\mu S$ and $\overline{H} \gamma^\mu \gamma_5 S$ are redundant since $\overline{H} \gamma^\mu S = \overline{H} \frac{1}{2}
\{\gamma^\mu, \rlap/v \}S = v^\mu\overline{H} S$ and $\overline{H} \sigma^{\mu \nu} S = \overline{H}\frac{1}{2} \{\sigma^{\mu \nu},
\rlap/v \}S = -\epsilon_{\mu \nu \alpha \beta}  v^\alpha \overline{H}\gamma^\beta \gamma_5 S$, while $\overline{H} \gamma_5 S = 0$.
Spin-symmetry violating operators will be of the form ${\rm Tr} [ \overline{H}\sigma^{\mu \nu} S \Gamma]$. $\Gamma$ must be
$\gamma^\alpha$ or $ \gamma_5$ since the trace vanishes for $\Gamma=1$, while $\Gamma=\gamma^\alpha \gamma_5$ and $\sigma^{\alpha \beta}$
are redundant because 
\bea 
{\rm Tr} [ \overline{H}\sigma^{\mu \nu} S \gamma^\alpha \gamma_5] &=&
v^\alpha {\rm Tr} [ \overline{H}\sigma^{\mu \nu} S \gamma_5]  \nn \\
{\rm Tr} [ \overline{H}\sigma^{\mu \nu} S \sigma^{\alpha \beta}] &=&
i \,{\rm Tr} [ \overline{H}\sigma^{\mu \nu} S (v^\alpha \gamma^\beta - v^\beta \gamma^\alpha ) ]   \, . \nn
\eea
Reparametrization invariance \cite{Luke:1992cs}
forbids operators with derivatives acting on the $H$ or $S$ fields~\cite{Falk:1992cx}.
For $S \to H \gamma$ decays, the lowest dimension, parity conserving, spin-symmetry violating operators are
\bea\label{nlo}
{\cal L} =  \frac{i e e_Q \tilde \beta^\prime}{8 m_Q} {\rm Tr} [ \overline{H}_a \sigma^{\mu \nu} S_a  \gamma_5] F^{\alpha \beta} 
\epsilon_{\mu \nu \alpha \beta} + \frac{e e_Q \tilde \beta^{\prime \prime}}{8 m_Q} 
{\rm Tr}[\overline{H}_a\sigma^{\mu \nu}S_a \gamma^\alpha] i \partial_\alpha F_{\mu \nu} \, + {\rm h.c.} \, .
\eea
The $1/m_Q$ dependence (expected for any operator which violates heavy-quark spin symmetry) is explicit. The factors of $i$ are  required
by time reversal invariance. The first operator in Eq.~(\ref{nlo}) and the leading operator in Eq.~(\ref{emlo}) have mass dimension 5. 
The second operator in Eq.~(\ref{nlo}) has mass dimension 6, so $\tilde \beta^{\prime \prime}$ has mass dimension $-1$ and is expected to
scale like  $1/\Lambda_\chi \sim {\rm GeV}^{-1}$.  Since $2 \sigma^{\mu \nu} \partial_\mu F_{\alpha \nu} = \sigma^{\mu \nu}
\partial_\alpha F_{\mu \nu}$ for an abelian field strength there is a unique way of contracting indices in this operator. Note that
there is a unique dimension 6, spin-symmetry conserving operator proportional  to  ${\rm Tr}[\overline{H}_a S_b \sigma^{\mu \nu}
Q^\xi_{ba}] i v \cdot \partial F_{\mu \nu}$. This operator gives slight deviations from the ratios in Eq.~(\ref{ratio}) since its
contribution is suppressed by $|\vec k_\gamma|/\Lambda_\chi$ and $\vk$ differs for the  three decays due to hyperfine
splittings. These corrections should be smaller than corrections coming from operators in Eq.~(\ref{nlo}) so they are neglected
in what follows.

Power counting is  used to determine the importance of higher dimension operators in the Lagrangian.
HH$\chi$PT is a double expansion in $\Lambda_{\rm QCD}/m_Q$ and $Q/\Lambda_{\rm \chi}$, where $Q \sim p \sim m_\pi \sim m_K$.
Two additional mass scales appearing in the Lagrangian of Eq.~(\ref{Lag}) are  the mass splitting between
the $H$ and $S$ doublet fields, $\delta_{SH}$, and the hyperfine splittings
within each doublet. In the heavy-quark limit, the $S$ field propagator is proportional to
\bea
\frac{i}{2 (v\cdot k -\delta_{SH})} \, .
\eea
If $\delta_{SH}$ were to scale as $Q^0$, then the $S$ propagator could be expanded in powers of $v \cdot k/\delta_{SH}$ since 
$v\cdot k \sim Q$. In the strange sector loops receive important contributions from momenta 
$\sim m_K = 495 \, {\rm MeV}$.  Numerically, $\delta_{SH} \approx 350 \, {\rm MeV}$ in the strange quark sector
and $\approx 430 \, {\rm MeV}$ in the nonstrange sector, so expanding in $v\cdot k/\delta_{SH}$
is a poor approximation. Therefore, $\delta_{SH} \sim Q$ is required.
The  hyperfine splittings are also treated as $\sim Q$ since numerically these splittings are $\approx 140 \, {\rm
MeV}$ which is $\sim m_\pi$. 

There are $SU(3)$ violating corrections to the decay rates from operators such as 
${\rm Tr}[\overline{H}_a S_b \sigma^{\mu\nu}] F_{\mu \nu} Q^\xi_{bc} m_{ca}^\xi$.
These operators will give the same correction to all three electromagnetic decays in Eqs.~(\ref{belle_one}-\ref{cleo}),
so their effect can be absorbed into the definition of $\tilde \beta$. However, if one were interested in relating the 
electromagnetic decays of strange and non-strange heavy mesons these operators must be included explicitly.

The leading operator in Eq.~(\ref{emlo})  is order $Q$ because of the derivative in $F_{\mu \nu}$.  The first
operator in Eq.~(\ref{nlo}) is treated as $\sim Q^2$ because it is suppressed by $\Lambda_{\rm QCD}/m_Q$ relative to
the leading operator. The second operator  in Eq.~(\ref{nlo}) has two derivatives and is also  $1/m_Q$ suppressed.
It is treated as $\sim Q^2$. The correctness of this power counting  is confirmed by the
calculation of one-loop corrections to the decay, since in order to properly renormalize these diagrams both
counterterms in Eq.~(\ref{nlo}) are needed. In loop diagrams, integrals scale as
$Q^4$, the propagators of the  $H$ and $S$ fields as $\sim Q^{-1}$ and the propagators of Goldstone bosons as $\sim
Q^{-2}$. The leading couplings of the $H$ and $S$ fields to kaons and pions are $\sim Q$ and the couplings of the photon
to the kaons and pions  are $\sim Q$.  Calculations of the loop corrections 
are performed in $v\cdot A =0$ gauge where the  leading coupling of the photon
to the heavy meson fields vanishes. Finally, there is an $\sim Q^0$ coupling of two heavy meson fields to a Goldstone
boson and photon which comes from gauging the derivative couplings of the heavy meson fields to Goldstone bosons.
The $H S \pi \gamma$ vertex obtained by gauging the  derivative on the pion field in the leading $H S \pi$ coupling
vanishes in $v\cdot A =0$ gauge. With these power counting rules, the loop diagrams shown in Fig.1 give an $O(Q^2)$
contribution to the  $S \to H \gamma$ decays. Double lines are $S$ fields, solid lines  are $H$ fields and the
dotted lines are Goldstone bosons. For $D_s$ decays the Goldstone bosons  in these loops are $K^+$ and the virtual
heavy mesons are neutral $D$'s. 

\begin{figure}[!t]
  \centerline{\epsfysize=7.0truecm \epsfbox[100 460 470 665]{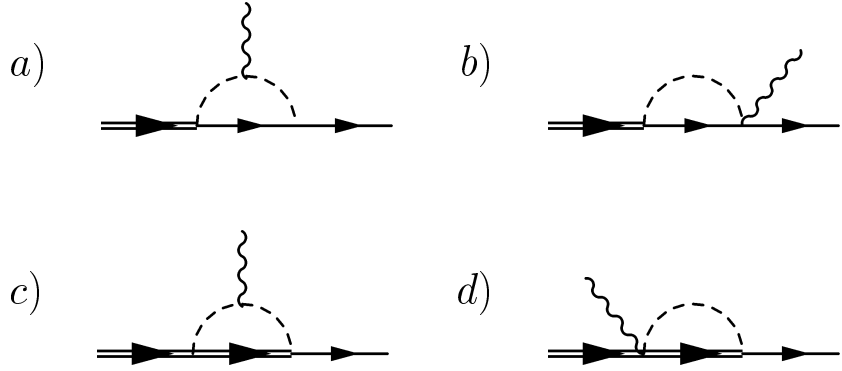}  }
 {\tighten
\caption[1]{One-loop chiral corrections to the electromagnetic decays
$S \to H \gamma$ in $v\cdot A=0$ gauge. Double lines are $S$ mesons,
solid lines are $H$ mesons, dashed lines are Goldstone bosons and 
the wavy line is the photon.}
\label{em} }
\end{figure}
Including both the loop diagrams and  tree level insertions of the operators in Eq.~(\ref{nlo}) 
yields:
\begin{eqnarray}\label{result}\hat 
\Gamma[1^+_a \to 1^-_a \gamma] &=& 1+\frac{2}{ e_q \tilde \beta} \left( - \frac{e_Q \tilde \beta^\prime}{m_Q} 
+ F\left[m_{1^+_a}-m_{1^-_b}  \, , m_{1^+_a}-m_{1^+_b} \, , \vk, M ,\mu \right]
\right)   \\
 \hat \Gamma[ 1^+_a \to 0^-_a  \gamma] &=& 1+\frac{2}{ e_q \tilde \beta} \left( \frac{e_Q \tilde \beta^\prime}{m_Q} 
 + \frac{e_Q \tilde \beta^{\prime \prime} \vk}{2 m_Q} +
 F\left[m_{1^+_a}-m_{1^-_b}  \, , m_{1^+_a}-m_{0^+_b}  \, , \vk, M,\mu \right] 
 \right)  \nn \\
\hat \Gamma[0^+_a \to 1^-_a \gamma] &=& 1+\frac{2}{ e_q \tilde \beta} \left(  \frac{e_Q \tilde \beta^\prime}{m_Q} 
- \frac{e_Q \tilde \beta^{\prime \prime} \vk}{2 m_Q} 
+ F\left[m_{0^+_a}-m_{0^-_b}  \, , m_{0^+_a}-m_{1^+_b}  \, , \vk, M,\mu \right]
 \right) \,. \nn 
\end{eqnarray}
Here $\hat \Gamma =\Gamma^{NLO}/\Gamma^{LO}$, where the 
$\Gamma^{LO}$ are given in Eq.~(\ref{lo}). The $SU(3)$ index $a$ refers 
to the external heavy mesons while the index $b$ refers to the mesons inside the
loop. $M$ is the mass of the virtual Goldstone boson. 
For  heavy-strange decays the external particles are heavy-strange mesons; $a=3$,
the Goldstone boson is a $K^+$, and the heavy-mesons inside the loops
are neutral heavy mesons with $b=1$. $\hat \Gamma$ is expanded to $O(Q)$. 
The function $F[\Delta_1,\Delta_2,\vk,M,\mu]$ is  given in the
Appendix. The loop graphs are regulated in dimensional regularization, counterterms are 
defined in the $\overline{MS}$ scheme and the dimensional regularization parameter is $\mu$. All 
 $\mu$ dependence is canceled by the implicit $\mu$ dependence of the renormalized
couplings $\tilde \beta$, $\tilde \beta^\prime$, and $\tilde \beta^{\prime \prime}$.  
 
An NLO calculation of the electromagnetic branching ratios also requires  $O(1/m_c)$
corrections to the decays $\Gamma[D_{s0}(2317)\to D_s \pi^0] $ and $\Gamma[D_{s1}(2460)\to D^*_s \pi^0]$.
The leading spin-symmetry violating operator contributing to these decays is
\bea
{\cal L} = \frac{h^\prime}{2 m_Q} {\rm Tr}[\overline{H}_a \sigma^{\mu \nu} S_b \gamma^\alpha] A_{b a}^\beta \epsilon_{\mu \nu \alpha \beta} \, .
\eea
Because of the $1/m_Q$ suppression this operator is considered $O(Q^2)$. The
one-loop diagrams contributing to $S \to H \pi$ transitions are subleading at $O(Q^3)$.
 The  decay rates to NLO are 
\bea\label{nlop}
\Gamma[1_3^+\to  1_3^- \pi^0] &=&\left(h - \frac{h^\prime}{m_Q}\right)^2
\frac{\theta^2}{3 \pi f^2} \frac{m_{1_3^-}}{m_{1_3^+}} E^2_{\pi^0} \vpi  \nn \\ 
\Gamma[0_3^+\to  0_3^-  \pi^0] &=&\left(h + 3 \frac{h^\prime}{m_Q}\right)^2
\frac{ \theta^2}{3 \pi f^2} \frac{m_{0_3^-}}{m_{0_3^+}} E^2_{\pi^0} \vpi    \, .
\eea
Earlier in this section the data from $D_0^0$ and $D_1^0$ decays was averaged 
to extract $h^2$. Including the leading
correction it is possible to fit $h$ and $h^\prime$ separately,
extracting $h=0.69\pm0.09$ and $h^\prime/m_c = -0.019 \pm 0.034$.

The NLO expression for the branching fraction ratios of heavy-strange mesons
is obtained by combining  Eq.~(\ref{nlop}) with Eq.~(\ref{result}) and Eq.~(\ref{lo}). The result is
\bea\label{answer}
\frac{{\rm Br}[1^+_3\to 1^-_3 \gamma]}{ {\rm Br}[1^+_3\to 1^-_3 \pi^0]}
&=& \frac{2\pi \alpha f^2 \tilde \beta^2}{9 \,\theta^2 h^2} \frac{\vk^3}{E_{\pi^0}^2 p_{\pi^0}} \times \nn\\
&&\hspace{-1.0 in}\left(1+\frac{2 h^\prime}{h \, m_Q} + \frac{6 e_Q \tilde \beta^\prime}{\tilde \beta \,m_Q} 
-\frac{6}{\tilde \beta} F\left[m_{1^+_a}-m_{1^-_b}  \, , m_{1^+_a}-m_{1^+_b} \, , \vk,m_{K^+} ,\mu \right]  \right) \nn \\ \nn \\
\frac{{\rm Br}[1^+_3 \to 0^-_3 \gamma]}{ {\rm Br}[1^+_3\to 1^-_3 \pi^0]}
&=& \frac{\pi \alpha f^2 \tilde \beta^2}{9\, \theta^2 h^2} \frac{m_{0^-_3}}{m_{1^-_3}}\frac{\vk^3}{E_{\pi^0}^2 p_{\pi^0}} \times \nn \\
&& \hspace{-1.0 in}\left(1+\frac{2 h^\prime}{h\, m_Q}-\frac{6 e_Q \tilde \beta^\prime}{\tilde \beta \,m_Q}  \ 
 - \frac{3 e_Q \tilde \beta^{\prime \prime} \vk}{ \tilde\beta \,m_Q} 
-\frac{6}{ \tilde \beta} F\left[m_{1^+_a}-m_{1^-_b}  \, , m_{1^+_a}-m_{0^+_b}  \, , \vk,m_{K^+} ,\mu \right] \right)\nn \\ \nn \\
\frac{{\rm Br}[0^+_3\to 1^-_3 \gamma]}{ {\rm Br}[0^+_3\to 0^-_3 \pi^0]}
&=& \frac{\pi \alpha f^2 \tilde \beta^2}{3\, \theta^2 h^2} \frac{m_{1^-_3}}{m_{0^-_3}}\frac{\vk^3}{E_{\pi^0}^2 p_{\pi^0}}\times   \\
&&\hspace{-1.0 in}\left(1-\frac{6 h^\prime}{h\, m_Q}-\frac{6 e_Q \tilde \beta^\prime}{ \tilde \beta \,m_Q} 
+ \frac{3 e_Q \tilde \beta^{\prime \prime} \vk}{ \tilde \beta \,m_Q} 
- \frac{6}{\tilde \beta}F\left[m_{0^+_a}-m_{0^-_b}  \, , m_{0^+_a}-m_{1^+_b}  \, , \vk,m_{K^+}   ,\mu \right]\right) \, , \nn
\eea
where $e_q$ has been set to $e_s= -1/3$. Applying the formulae in Eq.~(\ref{answer}) to the experimentally observed ratios 
gives
\begin{eqnarray}\label{csbr}
\frac{ {\rm Br}(D_{s1}(2460)\to D^*_s \gamma)}{{\rm Br}(D_{s1}(2460)\to D^*_s \pi^0)}  &=& 1.58 \, \frac{{\tilde \beta}^2}{h^2} \times \nn \\
&&\hspace{-0.55 in}\left(1 + \frac{1.43\, h^\prime}{h} 
+ \frac{2.86\, \tilde \beta^\prime}{\tilde \beta} 
+ \frac{0.18\, g \, h}{\tilde \beta} - \frac{2.94^{+0.70}_{-0.58}\, g^\prime \, h}{\tilde \beta} \right) \,  < \, 0.16  \nn \\ \nn \\
\frac{ {\rm Br}(D_{s1}(2460)\to D_s \gamma)}{{\rm Br}(D_{s1}(2460)\to D^*_s \pi^0)}  &=& 1.90 \, \frac{{\tilde \beta}^2}{h^2} \times\nn \\
&&\hspace{-1.25 in}\left(1 + \frac{1.43 \,h^\prime}{h} 
 - \frac{2.86 \,\tilde \beta^\prime}{\tilde \beta} - 
\frac{0.63\, \tilde \beta^{\prime \prime}}{\tilde \beta}  - \frac{0.03 \,g \, h}{\tilde \beta} 
- \frac{2.40^{+0.73}_{-0.59} \,g^\prime \, h}{\tilde \beta}\right) 
\, = \, 0.44 \pm 0.09   \nn \\ \nn \\
\frac{ {\rm Br}(D_{s0}(2317)\to D_s \gamma)}{{\rm Br}(D_{s0}(2317)\to D_s \pi^0)}  &=& 0.57 \, \frac{{\tilde \beta}^2}{h^2} \times \nn \\
&&\hspace{-1.25 in}\left(1- \frac{4.29 \,h^\prime}{h} -
 \frac{2.86 \,\tilde \beta^\prime}{\tilde \beta} - \frac{0.28\, \tilde \beta^{\prime \prime}}{\tilde \beta}
+ \frac{0.37\, g \, h}{\tilde \beta} - \frac{3.81^{+0.90}_{-0.75}\, g^\prime \, h}{\tilde \beta} \right) 
\, < \, 0.059 \, .
\end{eqnarray}
Here $\tilde \beta$ and $\tilde \beta^{\prime \prime}$ are measured in units of ${\rm GeV}^{-1}$  and $h^\prime$ in units of GeV. All
other quantities are dimensionless. The charm quark mass is $m_c = 1.4$ GeV, the renormalization scale  is $\mu = 1$ GeV, and  $e_Q = e_c
= 2/3$. For the loop corrections with kaons the meson decay constant is $f=f_K$, while for the strong decays  $f=f_\eta$. (If $f_\pi$ is
used  in the strong decays, then  the branching fraction ratios in Eq.~(\ref{answer}) should be multiplied by $f_\pi^2/f_\eta^2 = 0.58$.)
The masses used for the virtual nonstrange even-parity heavy mesons in the loops are $m_{0^+_1} = 2308 \pm 36$ MeV  and $m_{1^+_1} = 2438
\pm 29$ MeV, where the first number is the nonstrange $0^+$ mass measured by Belle and the second is the average of the  nonstrange $1^+$
mass measured by CLEO and Belle. The uncertainty in the coefficient of $g^\prime h/\tilde \beta$ in Eq.~(\ref{csbr}) is due to the uncertainty in
the masses of the $D_0^0$ and $D_1^0$. 

The result depends on seven parameters: $g,g^\prime, h, h^\prime, \tilde \beta, \tilde \beta^\prime$, and $\tilde \beta^{\prime \prime}$. 
The coupling $g$ is constrained to be $0.27^{+.06}_{-.03}$ from a next-to-leading order HH$\chi$PT 
analysis of $D^*$ decays~\cite{Stewart:1998ke}.  $h$ and $h^\prime$ are extracted from the nonstrange
decays, leaving four unknown parameters. Since there are only three constraints coming from experiment,
further analysis requires additional assumptions to constrain the parameter space.

To illustrate how the current data is consistent with natural size parameters the following situation is considered.
The contribution from $\tilde \beta^{\prime \prime}$ is neglected since
in Eq.~(\ref{csbr}) $\tilde \beta^{\prime \prime}$ is multiplied by a coefficient 
that is much smaller than the coefficients multiplying $h^\prime$ and $\tilde \beta^\prime$.
(The smallness of this coefficient is due to the factor $\vk/m_c$.) $g$, $h$, $h^\prime$, and the 
branching fraction ratio measured by Belle are set to their central values:
0.27, 0.69, -0.019 $m_c$, and 0.44, respectively. Ranges for the 
remaining parameters ($\tilde \beta,\tilde \beta^\prime$, and $g^\prime$) are extracted by varying the branching ratios 
in Eq.~(\ref{cleo}) between 0 and their upper limits. 
 There are two solutions since the formulae for the electromagnetic decay rate
is quadratic in $\tilde \beta$. The results are
\bea\label{ex_one}
0.70 \leq \tilde \beta \,{\rm GeV} \leq 0.86 \qquad -0.01 \leq &\tilde \beta^\prime& \leq 0.01 \qquad 0.32 \leq g^\prime \leq 0.40 \nn\\
\qquad \quad -0.62 \leq \tilde \beta \, {\rm GeV} \leq -0.46 \,\,\,\,\quad -0.01 \leq &\tilde \beta^\prime& \leq 0.02 \quad  
-0.25 \leq g^\prime \leq -0.16 \, .
\eea
Note that the ranges quoted in Eq.~(\ref{ex_one}) do not include errors due to the uncertainties in  the parameters $g$, $h$, and  $h^\prime$ or  the masses of
the $D_0^0$ and $D_1^0$. $h^\prime$ is highly uncertain because of the uncertainty in the  masses and widths of the $D_0^0$ and $D_1^0$ used to extract it. The
loop contribution proportional to $g^\prime h/\tilde \beta$ is also sensitive to the masses of the $D_0^0$ and $D_1^0$ that appear as intermediate states. The
ranges given in Eq.~(\ref{ex_one}) do not reflect these uncertainties and do not exhaust the possible parameter space. Instead, they are simply illustrative of
natural size parameters consistent with existing data. 

When more data on excited heavy meson systems becomes available, the formulae in Eq.~(\ref{answer})
could be used to test  models  that make predictions for the
parameters in HH$\chi$PT. In parity doubling models, $g = - g^\prime$ and $h=1$ at tree  level~\cite{Bardeen:2003kt}. The authors of
Ref.~\cite{Bardeen:2003kt} note that $h$ can be renormalized away from its tree level value and allow this parameter to vary in their
analysis of  strong decays. The tree level result $h=1$ exceeds the value extracted from excited nonstrange decays in HH$\chi$PT.  Another theoretical
framework which makes similar predictions for the coupling constants $g$, $g^\prime$, and $h$ is the algebraic realization of chiral
symmetry~\cite{Weinberg:xn}. Applying this theory to heavy mesons~\cite{Beane:2002td,Beane:1995yw} leads to the predictions $g^\prime = -g$ and $g^2 + h^2=1$.
Using $g=0.27$ in this relation gives $h^2= 0.93$ which is also larger than extracted from Eq.~(\ref{nlop}).
While the predictions for $h$ are not in agreement with available data, the condition $g =- g^\prime$ is consistent with available data but not
required.

Eventually the even parity $B_s$ states will be observed and all  electromagnetic branching fractions for  heavy-strange mesons will be
measured.  Then the parameter space will be overconstrained  and HH$\chi$PT for excited heavy mesons can be tested  decisively.
Furthermore,  the extracted values for $g, g^\prime$, and $h$ can be compared with predictions from parity doubling models and algebraic
realizations  of chiral symmetry.  At the present time, observed  violations of leading heavy-quark spin symmetry predictions are
consistent with what is expected from loop effects and  higher order operators appearing in the HH$\chi$PT Lagrangian.

\section{Electromagnetic Decays and D  K molecules}

The unexpectedly low masses of the $D_{s0}(2317)$ and $D_{s1}(2460)$ have prompted speculation that these states are unconventional. Two
common proposals are that these mesons are  $c \bar s q \bar q$  tetraquarks or hadronic bound states of $D$ and $K$ mesons. This section
addresses the question of what the decays reveal about the internal structure of the $D_{s0}(2317)$ and $D_{s1}(2460)$. In the
analysis  of the previous section the only information about the states needed  to construct the  HH$\chi$PT Lagrangian is the assumed
$SU(3)$ and $j^p$ quantum numbers of the light degrees of freedom in the hadrons. A constituent quark in a $P$-wave or an exotic with two
light quarks and an  antiquark both have $j^p =\frac{1}{2}^-$. Both states are represented by a field like that in Eq.~(\ref{S}).
Analysis of electromagnetic and strong decays within HH$\chi$PT is identical for both states, though the coupling constants $
\tilde \beta, \tilde \beta^\prime, h$, etc. would be different for the two states. Since these coupling constants are unknown in either case, the HH$\chi$PT
predictions for electromagnetic  and strong decays cannot distinguish between exotic $c \bar s q \bar q$ and conventional  $c\bar s$
P-wave states. Of course, if the $D_{s0}(2317)$ and $D_{s1}(2460)$ are $c\bar s q \bar q $ states then in the quark model there should be
distinct $c\bar s$ P-wave mesons with the same quantum numbers.  These states could be very hard to detect, however, if they are above
the $D K$ threshold. Mixing between the conventional and exotic mesons is also likely~\cite{Browder:2003fk,Nussinov:2003uj}.

However, if the $D_{s0}(2317)$ $(D_{s1}(2460))$ is a bound state of $D^{(*)}$ and $K$ mesons then the HH$\chi$PT predictions for
electromagnetic and strong decays will be different. For a hadronic bound state of a $D$ or $D^*$ and a kaon, one could in principle
calculate the bound state masses and  other properties from the HH$\chi$PT Lagrangian with the field $H_a$ alone. There have been
attempts to generate the $D_{s0}(2317)$ and $D_{s1}(2460)$ as resonances in a unitarized meson
model~\cite{vanBeveren:2003kd,vanBeveren:2003hj} as well as by solving Bethe-Salpeter equations in relativistic, unitarized chiral
perturbation theory~\cite{Kolomeitsev:2003ac,Hofmann:2003je,Lutz:2004tv}. Producing a bound state requires resumming an infinite number
of Feynman graphs in HH$\chi$PT and  the renormalization of these graphs requires introducing higher order operators whose renormalized 
coefficients  are unknown. Such a calculation will not be attempted in this paper. Instead the $D K$  molecular picture will be tested by
simply assuming that strong forces between $D^{(*)}$  and $K$ mesons give rise to  the $D_{s0}(2317)$ and $D_{s1}(2460)$ and determining
what this implies for the  decay rates. If the $D_{s0}(2317)$ ($D_{s1}(2460)$) are bound states of $D^{(*)} K$ then the characteristic
momentum of the constituents  is  $p \sim \sqrt{2 \mu B} \approx 190$ MeV, where $\mu$ is the  reduced mass and $B$ is the binding
energy. The $D K$ molecule can then be modeled as a nonrelativistic bound state since  relativistic corrections are suppressed by $v^2=
p^2/M_K^2 = 0.15$. Strong and electromagnetic decays can be calculated in terms of the unknown bound state wavefunction. Even without any
knowledge of these wavefunctions it is   possible to  make predictions for the decay ratios in Eqs.~(\ref{belle_one}-\ref{cleo}). It
turns out that these  predictions disagree with data so interpreting $D_{s0}(2317)$ and $D_{s1}(2460)$ as $D K$  molecules is disfavored.

Ref.~\cite{Godfrey:2003kg} advocates using the radiative decays of the $D_{s0}(2317)$ and $D_{s1}(2460)$  to determine the nature of
these states and calculates radiative and strong decays within a non-relativistic quark model. Predictions for the branching fraction
ratios in Eqs.~(\ref{belle_one}-\ref{cleo})  are in the same proportion as leading order heavy-quark symmetry predictions, though they
are approximately 45\% larger than the leading order predictions obtained in section II. The quark model expectation for the total widths
of the $D_{s0}(2317)$ and $D_{s1}(2460)$ is $O(10 \,{\rm keV})$, consistent with Eq.~(\ref{singpi}). However, the  conclusion of
Ref.~\cite{Godfrey:2003kg}  states that a $D^{(*)}K$ molecule  should have a width of $O(1\, {\rm MeV})$ and that the electromagnetic
transitions should be absent. The analysis that follows is consistent with the first conclusion but not the second. Below it is
demonstrated that the electromagnetic branching ratios of a $D^{(*)}K$ molecule are large and are in worse agreement with experiment than
the nonrelativistic quark model.

\begin{figure}[!t]
  \centerline{\epsfysize=8.0truecm \epsfbox[90 490 480 730]{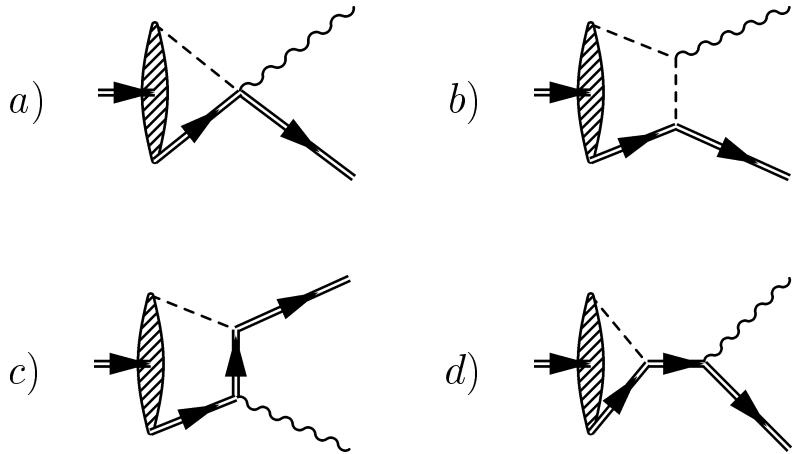}  }
 {\tighten
\caption[1]{Leading order diagrams for $D^{(*)}K$ bound states decaying 
into $D_s^{(*)} \gamma$. The shaded oval represents the $D^{(*)} K$ 
bound state wavefunction. }
\label{DK_photo} }
\end{figure}

The $D_{s0}(2317)$ ($D_{s1}(2460)$) is assumed to be an $S$-wave $I=0$ bound state of $D^{(*)}$ and $K$ mesons.
The matrix elements for the electromagnetic decays of the $D_{s0}(2317)$ and $D_{s1}(2460)$ are given by:
\bea\label{mas}
{\cal M}[D_{s0}(2317) \to D_s^* \gamma] &=&\sqrt{\frac{2}{m_{D_{s0}}}}\int\frac{d^3 {\bf p}}{(2\pi)^3}\, 
\tilde \psi_{DK}(\vec{p})\, {\cal M}[D(\vec{p}) K(-\vec{p}) \to D_s \gamma] \nn \\
{\cal M}[D_{s1}(2460) \to D_s^{(*)} \gamma] &=&\sqrt{\frac{2}{m_{D_{s1}}}}\int\frac{d^3 {\bf p}}{(2\pi)^3} \,
\tilde\psi_{D^*K}(\vec{p})\, {\cal M}[D^*(\vec{p}) K(-\vec{p}) \to D^{(*)}_s \gamma] \, .
\eea
Here ${\bf p}$ is the three-momentum of the $D^{(*)}$ meson in the bound state and 
$\tilde\psi_{D^{(*)}K}(\vec{p})$ is the bound state  momentum-space wavefunction. 
Although calculation of the bound state wavefunction is nonperturbative, the 
typical momentum is small enough that the amplitudes ${\cal M}[D^{(*)}  K  \to D^{(*)}_s \gamma]$ are perturbatively calculable in 
HH$\chi$PT.  The leading order diagrams for the decay rates in Eq.~(\ref{mas}) are shown in Fig.~\ref{DK_photo}. 
The shaded oval on the left hand side of these Feynman diagrams represents  the $D^{(*)} K$ molecule. In Figs.~\ref{DK_photo}a-b the dashed line is a $K^+$,
and the vertex involving the photon comes from gauging the  $D^*$-$D_s$-$K^+$ coupling (Fig.~\ref{DK_photo}a) or the $K^+$ kinetic term
(Fig.~\ref{DK_photo}b). In Figs.~\ref{DK_photo}c-d, the photon coupling comes from gauging the heavy meson kinetic term. There are also 
diagrams  like Figs.~\ref{DK_photo}c-d where the photon heavy-meson coupling  comes from a term in the Lagrangian proportional to  ${\rm
Tr}[\overline H_b H_a \sigma^{\mu \nu} Q^\xi_{a b}] \,F_{\mu \nu}$, but these only contribute in the $P$-wave channel.

The graphs in Fig.~\ref{DK_photo}c-d are nonvanishing in the $K^0 D^+$ channel, but are equal and 
opposite in sign so the contribution in this channel vanishes. The graph in  Fig.~\ref{DK_photo}c vanishes in the $K^+ D^0$ channel. 
The amplitudes are  
\bea\label{sd}
{\cal M} [D^*(\vec{p}) K(-\vec{p}) \to D^*_s \gamma] &=& - i\sqrt{m_{D^*} m_{D^*_s}}
\frac{2 e g}{f} \left[ \frac{p_K^\mu (p_K-p_\gamma)^\delta}{p_K\cdot p_\gamma}+ g^{\mu \delta}  
- \frac{p_K^\delta v^\mu}{v \cdot p_K}\right] \epsilon_{\mu}^*
\epsilon_{\nu \beta \lambda \delta} v^\beta   \epsilon_3^{* \nu} \epsilon_1^\lambda \nn \\
{\cal M} [D^*(\vec{p}) K(-\vec{p}) \to D_s \gamma] &=& \sqrt{m_{D^*} m_{D^*_s}}
\frac{2 e g}{f}\left[ \frac{p_K^\mu (p_K -p_\gamma)^\nu}{p_K\cdot p_\gamma}+g^{\mu \nu}
- \frac{v^\mu p_K^\nu}{v \cdot p_K}\right] \epsilon_{\mu}^*  (\epsilon_1)_\nu \nn \\
{\cal M} [D(\vec{p}) K(-\vec{p}) \to D^*_s \gamma] &=& \sqrt{m_{D} m_{D^*_s}}
\frac{2 e g}{f}\left[ \frac{p_K^\mu (p_K -p_\gamma)^\nu}{p_K\cdot p_\gamma}+g^{\mu \nu}
- \frac{v^\mu p_K^\nu}{v \cdot p_K}\right]  \epsilon_{\mu}^*  (\epsilon_3^*)_\nu \, .
\eea
Here $p_K$ and $p_\gamma$ are the kaon and photon four-momentum, respectively.  The polarization vectors for the photon, $D^*$,
and $D_s^*$ are denoted $\epsilon$, $\epsilon_1$, and $\epsilon_3$, respectively. It is  easy to check
that the amplitudes respect the QED Ward identity. These expressions are inserted into
Eq.~(\ref{mas}), $p_K^\mu$ is set to $E_K v^\mu + \vec{p}^\mu$, $\vec{p}^\mu = (0,-\vec{p})$, and the matrix element
is expanded to lowest order in $\vec{p}$.  Because of the rotational symmetry of the $S$-wave
wavefunction, $\tilde \psi(\vec{p})$, terms linear in $\vec{p}$ vanish.  There are corrections
to the amplitudes from higher orders in chiral perturbation theory that are $O(m_K^2/\Lambda_\chi^2)$
and relativistic corrections of $O(v^2)$. The errors in the predictions for the decay rates
could be as large as 50\%.

The results for the decay rates are  
\bea
\Gamma[D_{s1}(2460) \to D_s^* \gamma] &=& \frac{8 g^2 \alpha}{3 f^2}\left(\frac{m_{D^{0*}} m_{D_s^*}}{m_{D_{s1}}^3}\right)
|\psi_{D^* K}(0)|^2  \, \vk \nn \\
\Gamma[D_{s1}(2460) \to D_s \gamma] &=& \frac{4 g^2 \alpha}{3 f^2} \left(\frac{m_{D^{0*}} m_{D_s}}{m_{D_{s1}}^3}\right)
|\psi_{D^* K}(0)|^2  \vk \nn \\
\Gamma[D_{s0}(2317) \to D_s^* \gamma] &=& \frac{4 g^2 \alpha}{f^2}\left(\frac{m_{D^{0}} m_{D_s^*}}{m_{D_{s0}}^3}\right) 
|\psi_{D K}(0)|^2 \vk \, .
\eea
Here $\psi_{D K}(0) [\psi_{D^* K}(0)]$ is the wavefunction at the origin for the  $D_{s0}(2317) [D_{s1}(2460)]$. 
In the heavy-quark limit the   partial width ratios are again  $\Gamma[D_{s1}(2460) \to D_s^*
\gamma]:\Gamma[D_{s1}(2460) \to D_s \gamma]: \Gamma[D_{s0}(2317) \to D_s^* \gamma]=2:1:3$. However, the decay 
rates are proportional  to $\vk$ instead of $\vk^3$. This important
difference in the kinematic factors  leads to a very different prediction for the relative sizes of the
partial widths than obtained in Eq.~(\ref{ratio}). In this case
\bea
\Gamma[D_{s1}(2460) \to D_s^* \gamma]:\Gamma[D_{s1}(2460) \to D_s \gamma]:
\Gamma[D_{s0}(2317) \to D_s^* \gamma]=1.57:1:R_\psi 1.58 \, , \nn
\eea
where $R_\psi = |\psi_{D K}(0)|^2/|\psi_{D^* K}(0)|^2$ is expected to be $\approx 1$.
In this scenario $\Gamma[D_{s1}(2460) \to D_s \gamma]$ is the smallest decay rate rather than the largest.

\begin{figure}[!t]
  \centerline{\epsfysize=3.9truecm \epsfbox[0 600 550 720]{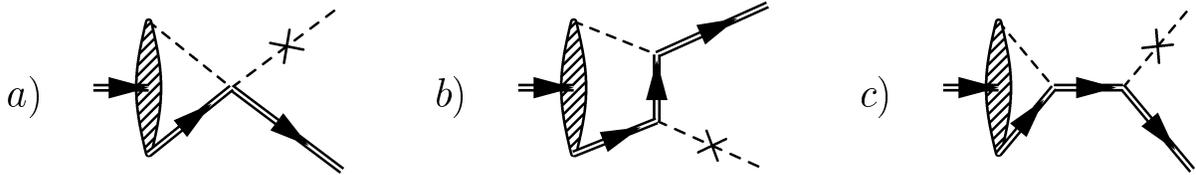}}
 {\tighten
\caption[1]{Leading order diagram for $D^{(*)}K$ bound states decaying 
into $D_s^{(*)} \pi^0$. 
The dashed line from the  bound state is a $K$, the dashed line 
in the final state is an $\eta$ which mixes into a $\pi^0$. }
\label{DK_strong} }
\end{figure}

To compare with the measured branching ratios the strong decays must also be calculated.
The leading order diagrams are shown in Fig.~\ref{DK_strong}. All three diagrams
depict $D^{(*)} K \to D_s^{(*)} \eta$ followed by $\eta-\pi^0$ mixing, which 
is represented by a cross on the dashed line in the final state. The vertex 
for the graph in Fig.~\ref{DK_strong}a comes from the chirally covariant 
derivative in the heavy meson kinetic term. This  graph  
contributes in the $S$-wave channel while  Figs.~\ref{DK_strong}b-c contribute to the $P$-wave channel only.
The results for the decay rates are
\bea \label{modelsucks}
\Gamma[D_{s1}(2460) \to D_s^* \pi^0] &=& 
\frac{3(m_K+E_{\pi^0})^2 \theta^2}{4 \pi f^4}\left(\frac{m_{D^*} m_{D^*_s}}{m_{D_{s1}}^3}\right) |\psi_{D^* K}(0)|^2 \vpi \nn \\
\Gamma[D_{s0}(2317) \to D_s  \pi^0] &=& 
\frac{3(m_K+E_{\pi^0})^2 \theta^2}{4 \pi f^4}\left(\frac{m_D m_{D_s}}{m_{D_{s0}}^3}\right) |\psi_{D K}(0)|^2 \vpi
\eea
If $f =f_\pi = 130$ MeV then the bounds $\Gamma[D_{s1}(2460)] \leq 7$ MeV and  $\Gamma[D_{s0}(2317)] \leq 7$ MeV
imply $|\psi_{D^{(*)} K}(0)|^2 \leq (52 \,{\rm MeV})^3$. If instead  $f =f_\eta = 171$ MeV then
$|\psi_{D^{(*)} K}(0)|^2 \leq (75 \,{\rm MeV})^3$. In either case the bounds on the wavefunctions are somewhat smaller
than expected: $|\psi_{D^{(*)} K}(0)|^2 \sim |{\bf p}|^3 \sim (190 \,{\rm MeV})^3$. Since this is only an order of magnitude estimate,
the bounds on $|\psi_{D^{(*)} K}(0)|^2$ are not a problem for the $D K$ molecular interpretation.
However, they do imply that if the $D_{s0}(2317)$ and $D_{s1}(2460)$  are $D^{(*)}K$ molecules the states should not be much narrower than the
present upper limits~\cite{Barnes:2003dj,Godfrey:2003kg}. 

Since the wavefunction squared cancels in the ratio of strong and electromagnetic decays 
the electromagnetic branching fractions can be predicted:
\bea
\frac{ {\rm Br}(D_{s1}(2460)\to D^*_s \gamma)}{{\rm Br}(D_{s1}(2460)\to D^*_s \pi^0)}  &=&  3.23   \nn \\
\frac{ {\rm Br}(D_{s1}(2460)\to D_s \gamma)}{{\rm Br}(D_{s1}(2460)\to D^*_s \pi^0)} &=&  2.21   \nn \\
\frac{ {\rm Br}(D_{s0}(2317)\to D_s \gamma)}{{\rm Br}(D_{s0}(2317)\to D_s \pi^0)}  &=&  2.96  \, .
\eea 
In this calculation  $\alpha =1/137$, $\theta= 0.01$, $g=0.27$,  $f_K = 159$ MeV in the electromagnetic decays 
and $f = f_\eta = 171$ MeV in the strong decays. If instead $f=f_\pi = 130$ MeV is used in the strong decays the 
predicted branching fraction ratios are smaller by a factor of three. While the branching fraction ratios 
are quite sensitive to the choice of $f$, in any case they  are much too
large compared to experiment. Also, the relative sizes of the branching fraction ratios are in disagreement with experiment,
since the second branching fraction ratio in Eq.~(\ref{modelsucks}) is predicted to be smallest, not largest. 
Note that the possibility of these states being mixtures of quark level bound states and $D K$
molecules~\cite{Browder:2003fk,Nussinov:2003uj} is also disfavored  since this would enhance the first and third
ratios in Eq.~(\ref{modelsucks}) relative to the second, whereas in reality these ratios are suppressed relative to the 
leading order  prediction in Eq.~(\ref{ratio}).

\section{Conclusions}

In this paper, corrections to electromagnetic and strong decays of  
$D_{s0}(2317)$ and $D_{s1}(2460)$ are calculated in HH$\chi$PT.
The corrections depend on a number of unknown or poorly determined coupling constants.
These predictions can be consistent with  Belle and CLEO data
with coupling constants of natural size.
Serious tests of the HH$\chi$PT description of the $D_{s0}(2317)$ and $D_{s1}(2460)$
will require more data on the electromagnetic branching ratios
of the even-parity charmed strange mesons and the strong decays of their nonstrange partners
as well as the decays of even-parity bottom strange mesons yet to be observed.
The work in this paper provides further stimulus for better experimental measurements
of charmed strange decays as well as discovery of their bottom strange counterparts.
Once better data becomes available, it would be interesting to test models of chiral symmetry
breaking which make specific predictions for the coupling constants appearing in the HH$\chi$PT 
lagrangian.

This paper  also  tests the hypothesis that the $D_{s0}(2317)$ and $D_{s1}(2460)$ are molecular 
bound states of $D K$ and $D^* K$ molecules, respectively. In this scenario,
these states are sufficiently nonrelativistic that HH$\chi$PT 
can be used to predict the decay rates at lowest order. Furthermore, bound state wavefunctions
cancel out of predictions for the observed branching fraction ratios so absolute
predictions can be made. These predictions are in much worse
agreement with data than leading order HH$\chi$PT predictions. Specifically,
predictions for all the  branching fraction ratios are larger than observed and  
the branching fraction for the only observed electromagnetic decay  is predicted to be the smallest
of the three possible decays rather than the largest. Therefore, a molecular interpretation of these states
is disfavored by available data on electromagnetic decays.

R.P.S. and T.M. are supported in part by DOE grant 
DE-FG02-96ER40945. T.M. is also supported in part by DOE grant DE-AC05-84ER40150.  
T.M. would  like to thank the Aspen Center for Physics where
part of this work was completed.

\section {Appendix}

The function $F[\Delta_1,\Delta_2,\vk,M,\mu]$ is
\bea
 F\left[\Delta_1,\Delta_2,\vk,M,\mu \right] &=& 
 \frac{g \,h}{8 \pi^2 f^2} \left[\Delta_1 \left(3 + {\rm ln}\left(\frac{\mu^2}{\Delta_1^2}\right)\right)+
 \frac{\Delta_1}{\vk^2} G\left[\Delta_1,\vk,M \right] \right] \\
&&+ \frac{g^\prime \,h}{8 \pi^2 f^2} \left[-(\Delta_2-\vk)  {\rm ln}\left(\frac{\mu^2}{\Delta_2^2}\right) +
\frac{(\Delta_2-\vk)}{\vk^2}   H\left[\Delta_2,\vk,M\right]\right] \, , \nn
\eea
where 
\bea
G(\Delta,\vk ,M) &=& -2 \vk \Delta - \vk^2 {\rm ln}\left(\frac{M^2}{\Delta^2}\right)
+ (\Delta - \vk)^2 F_1\left(\frac{\Delta -\vk}{M}\right) \nn \\
&& -\Delta (\Delta -2 \vk) F_1\left(\frac{\Delta}{M}\right)  +
M^2 \left[ F_2\left(\frac{\Delta}{M}\right) -  F_2\left(\frac{\Delta- \vk}{M}\right) \right] \, ,\nn
\eea
and
\bea
H(\Delta,\vk ,M) &=& -\vk^2-2 \vk \Delta  + \vk^2 {\rm ln}\left(\frac{M^2}{\Delta^2}\right)
+ (\Delta^2- \vk^2) F_1\left(\frac{\Delta -\vk}{M}\right) \nn \\
&& -\Delta^2 F_1\left(\frac{\Delta}{M}\right)  +
M^2 \left[ F_2\left(\frac{\Delta}{M}\right) -  F_2\left(\frac{\Delta- \vk}{M}\right) \right] \,.  \nn
\eea
The functions $F_{1,2}(x)$ are given by:
\bea
F_1(x) &=& \frac{2 \sqrt{1-x^2}}{x}\left[\frac{\pi}{2}-\arctan\left(\frac{x}{\sqrt{1-x^2}}\right)\right]
\qquad |x| <1 \\
&=&  -\frac{2 \sqrt{x^2-1}}{x} {\rm ln}\left(x+\sqrt{x^2-1}\right) \qquad \qquad \quad |x| >1 \nn \\
F_2(x) &=& \left[\frac{\pi}{2}-\arctan\left(\frac{x}{\sqrt{1-x^2}}\right)\right]^2 \nn \qquad \qquad \qquad |x| <1\\
&=& -{\rm ln}^2\left(x +\sqrt{x^2-1}\right) \qquad \qquad \qquad \qquad \quad |x| >1 \nn
\eea

\end{document}